\documentclass{article}
\usepackage{geometry}                
\usepackage{graphicx}
\usepackage{epstopdf}

\usepackage{url}
\usepackage{color}
\usepackage{amsmath}
\usepackage{mathtools}

\DeclarePairedDelimiter\ket{\lvert}{\rangle}
\DeclarePairedDelimiterX\braket[2]{\langle}{\rangle}{#1 \delimsize\vert #2}

\usepackage{amssymb}
\usepackage{ulem}

\usepackage[affil-it]{authblk}
\begin{document}

\title{Scattering theory of the bifurcation\\in quantum measurement}
\author{Karl-Erik Eriksson}
\author{Kristian Lindgren\footnote{Corresponding author: kristian.lindgren@chalmers.se}}

\affil{Complex systems group \\ Department of Space, Earth and Environment\\ 
Chalmers University of Technology, G\"oteborg, Sweden}


\date{\today}

\maketitle

\begin{abstract}
We model quantum measurement of a two-level system $\mu$. Previous obstacles for understanding the measurement process are removed by basing the analysis of the interaction between $\mu$ and the measurement device on quantum field theory. We show how microscopic details of the measurement device can influence the transition to a final state. A statistical analysis of the ensemble of initial states reveals that those initial states that are efficient in leading to a transition to a final state, result in either of the expected eigenstates for $\mu$, with probabilities that agree with the Born rule.\\
\end{abstract}


\noindent
Quantum mechanics is at the basis of all modern physics and fundamental for the understanding of the world that we live in. As a general theory, quantum mechanics should apply also to the measurement process. From the general experience of non-destructive measurements, we draw conclusions about the interaction between the observed system and the measurement apparatus and how this can be described within quantum mechanics.

We thus consider a quantum system $\mu$, interacting with a measurement device. For simplicity we assume that $\mu$ is a two-level system that is not destroyed in the process. Then after the measurement, $\mu$ ends up in one of the eigenstates of the measured observable. If $\mu$ is prepared in one of these eigenstates, it remains in that state after the measurement. If $\mu$ is initially in a superposition of the two eigenstates, it still ends up in one of the eigenstates and the measurement result is the corresponding eigenvalue. The probability for a certain outcome is the squared modulus of the corresponding state component in the superposition (Born's rule).

We have to show that these characteristics of the measurement process are consequences of deterministic quantum mechanics applied to the interaction between the system $\mu$ and the measurement apparatus. In a decoherence process, information would be lost and it would not lead to a pure final state for $\mu$.

Our idea is that the microscopic details of the measurement apparatus affect the process so that it takes $\mu$ into either of the eigenstates of the measured observable and initiates a recording of the corresponding measurement result. This bifurcation leading to one of the two possible final states for $\mu$ with a frequency given by Born's rule, has to be analyzed.

The requirement that $\mu$, if initially in an eigenstate of the observable, remains in that eigenstate after interacting with the apparatus, is usually considered to lead to a well-known dilemma: If applying the (linear) quantum mechanics of the 1930s to $\mu$ in an initial superposition of those eigenstates, the result of the process appears to be a superposition of the two possible resulting states for $\mu$ and the apparatus without any change in the proportions between the channels. This has been referred to as von Neumann's dilemma \cite{Nauenberg2011}.

Attempts to get around this problem include Everett's relative-state formulation \cite{Everett1957} and its continuation in DeWitt's many-worlds interpretation \cite{DeWitt1970} as well as non-linear modifications of quantum mechanics \cite{Gisin1984,Gisin1992,Percival1994,Ghirardi1986,Ghirardi1990}.

Bell pointed out that the Everett-DeWitt theory does not properly reflect the fact that the presence of inverse processes and interference are inherent features of quantum mechanics \cite{Bell2004}:

\begin{quote}
Thus DeWitt seems to share our idea that the fundamental concepts of the theory should be meaningful on a microscopic level and not only on some ill-defined macroscopic level. But at the microscopic level there is no such asymmetry in time as would be indicated by the existence of branching and the non-existence of debranching. [...] [I]t even seems reasonable to regard the coalescence of previously different branches, and the resulting interference phenomena, as the characteristic feature of quantum mechanics. In this respect an accurate picture, which does not have any tree-like character, is the 'sum over all possible paths' of Feynman.
\end{quote}

As suggested by Bell, we look into work of Feynman for a correct theory. We choose the scattering theory of quantum
field theory, including Feynman diagrams.

Since $\mu$ and the measurement apparatus first approach each other, then interact and after that separate, scattering theory should be adequate for describing the process. As will be seen, via inverse processes, scattering theory introduces the non-linearity that is necessary for avoiding von Neumann's dilemma.

{\it Consequences of scattering theory.}---The measurement device, or a sufficiently large part of the system that $\mu$ interacts with, will be denoted by $A$. Since we are dealing with a two-level system $\mu$, the Pauli matrices provide a suitable formalism with 
$\sigma_3=
\begin{psmallmatrix}
1 & 0 \\ 0 & -1
\end{psmallmatrix}$ 
representing the observable to be measured, with eigenstates 
$\ket{+}_\mu = 
\begin{psmallmatrix}
1 \\ 0
\end{psmallmatrix}$
and 
$\ket{-}_\mu =
\begin{psmallmatrix}
0 \\ 1
\end{psmallmatrix}$.

Let us investigate the characteristics of the interaction between $\mu$ and $A$ in scattering theory for the case with $A$ in a state with (unknown) microscopic details that are summarized in a variable $\alpha$. We then denote the normalized initial state of $A$ by $\ket{0,\alpha}_A$ (with $0$ indicating a state of preparedness). This means that we assume $\alpha$ to represent one microstate in an ensemble of possible initial states.

If $\mu$ is initially in the state $\ket{j}_\mu$ $(j = + \text{ or } -)$, after the interaction with $A$, its state remains the same, while $A$ changes from the initial state $\ket{0,\alpha}_A$ to a final state $\ket{j,\beta_j(\alpha)}_A$, also normalized. The first $j$ here indicates that $A$ has been marked by the state $\ket{j}_\mu$ of $\mu$. All other characteristics of the final state of $A$ are collected in $\beta_j(\alpha)$.

For a general normalized state of $\mu$, $\ket{\psi}_\mu = \psi_{+} \ket{+}_\mu + \psi_{-} \ket{-}_\mu$,
($| \psi_{+} |^2 + | \psi_{-} |^2 = 1$) the combined initial state of $\mu \cup A$ is
\begin{equation}
\label{initial state}
\ket{\psi}_\mu \otimes \ket{0,\alpha}_A = 
\big( \psi_{+} \ket{+}_\mu + \psi_{-} \ket{-}_\mu \big)
\otimes \ket{0,\alpha}_A \;.
\end{equation}
A measurement of $\sigma_3$ on $\mu$ leads to a certain result. Since two different results are possible, the $\mu A$-interaction should in general result in a transition to one of the following states,
\begin{equation}
\label{final states}
\ket{+}_\mu \otimes \ket{+,\beta_+(\alpha)}_A \; \text{,  or  } \; \ket{-}_\mu \otimes \ket{-,\beta_-(\alpha)}_A \;.
\end{equation}
The conclusion is then that the outcome must depend on the initial state of $A$, i.e., on $\alpha$.

\begin{figure}[htpb]
\begin{center}
\includegraphics[scale=.25]{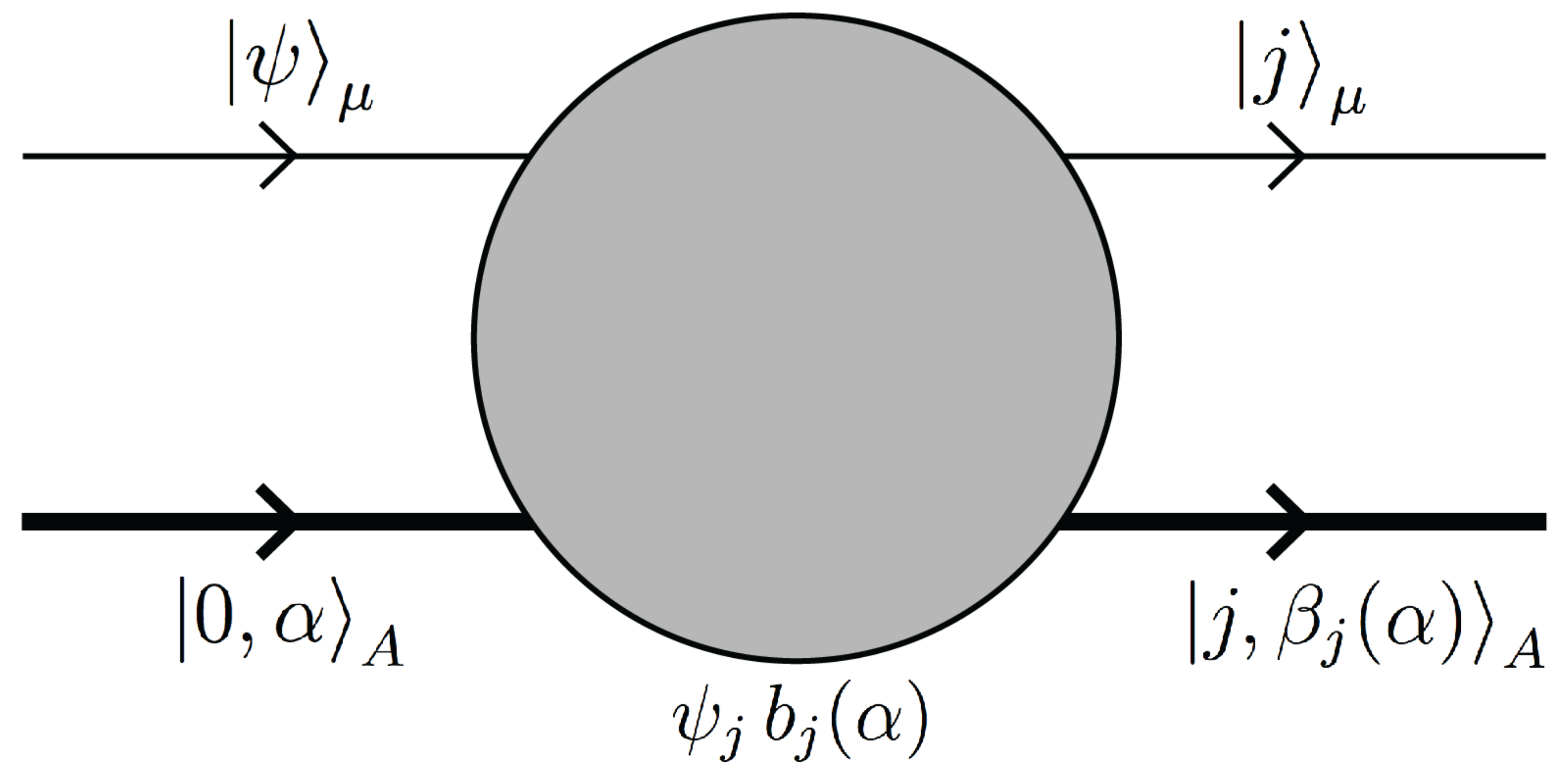}
\caption{Feynman diagram for a transition from the initial state $\ket{\psi}_\mu \otimes \ket{0,\alpha}_A$ to the final state $\ket{j}_\mu\otimes\ket{j,\beta_j(\alpha)}_A$, $j = \pm$. The transition amplitude $\psi_j b_j(\alpha)$ depends on the microscopic details of the initial state $\ket{0,\alpha}_A$ of the larger system $A$ and on the initial state $\ket{\psi}_\mu$ of $\mu$.}
\label{fig:Transition}
\end{center}
\end{figure}

In scattering theory, the interaction between $\mu$ and $A$ is characterised by the transition operator $M$, and this leads to the (non-normalized) final state (see Figure 1), 
\begin{equation}
\label{transitionM}
M \ket{\psi}_\mu \otimes \ket{0,\alpha}_A = 
b_+(\alpha)  \psi_{+} \ket{+}_\mu \otimes \ket{+,\beta_+(\alpha)}_A + b_-(\alpha) \psi_{-} \ket{-}_\mu \otimes \ket{-,\beta_-(\alpha)}_A  \;.
\end{equation}
In general, the amplitudes, $b_+(\alpha)$ and $b_-(\alpha)$, are not equal and therefore the proportions between $+$ and $-$ can change in a way that depends on the initial state $\ket{0,\alpha}_A$ of $A$. (Note that $M$ must not to be confused with the unitary scattering operator $S$; see the Supplemental Material \cite{supplementary}.)

The requirement of a statistically unbiased measurement means that $\langle\langle |b_+|^2 \rangle\rangle = \langle\langle |b_-|^2 \rangle\rangle$, where $\langle\langle \; \rangle\rangle$ denotes mean value over the ensemble of initial states $\ket{0,\alpha}_A$ of $A$.

Equation (\ref{transitionM}) describes a mechanism of the measurement process in which von Neumann's dilemma is not present. Since relativistic quantum mechanics, in the form of scattering theory of quantum field theory, is a more correct theory than the non-relativistic Schr\" odinger equation, as it was used in the 1930s, we choose to use Equation (\ref{transitionM}) as our starting point.

In quantum field theory, the two channels are connected via the initial state and inverse processes. A formulation based on perturbation theory with Feynman diagram representation to all orders, leads to an explicitly unitary description of the whole process. (This is shown in the Supplemental Material \cite{supplementary}.)

For equation (\ref{transitionM}) to properly represent a measurement process, i.e., a bifurcation that leads to a final state with $\mu$ in either of the eigenstates of $\sigma_3$, it is necessary that the squared moduli of the amplitudes satisfy either $|b_+|^2>>|b_-|^2$ or $|b_-|^2>>|b_+|^2$. If this holds for (almost) all microstates $\alpha$ in the resulting ensemble of final states, it can function as a mechanism for the bifurcation of the measurement process.

The von Neumann dilemma came from assuming $A$ to be in a given initial state $\ket{0,\alpha}_A$. For the initial state of $\mu A$-interaction, $A$ can be in any of the states of the available initial ensemble. These states are ready to influence the recording process in different ways. To reach a final state, given by Equation (\ref{transitionM}), they compete with their transition rates, $(2 \pi)^{-1} \big( |\psi_{+}|^2 |b_+(\alpha)|^2 + |\psi_{-}|^2 |b_-(\alpha)|^2 \big)$, which can differ widely between different values of $\alpha$. The competition leads to a selection and to a statistical distribution over $\alpha$ of the final states that is very different from the distribution in the initial ensemble.

It remains to be shown how the bifurcation leading to either $+$ or $-$ can occur, i.e., how we get either $|b_+|^2>>|b_-|^2$ or $|b_-|^2>>|b_+|^2$.

If this can be shown, however, we can see already now that because $\langle\langle |b_+|^2 \rangle\rangle = \langle\langle |b_-|^2 \rangle\rangle$, in the mean, the partial transition rates, $(2 \pi)^{-1} |\psi_{+}|^2 |b_+(\alpha)|^2$ and $(2 \pi)^{-1} |\psi_{-}|^2 |b_-(\alpha)|^2$, are proportional to $|\psi_{+}|^2$ and $|\psi_{-}|^2$, which thus are the probabilities for the final results $+$ or $-$. This means that under the stated condition, we have obtained Born's rule.

{\it Schematic mathematical model of a measurement device.}---Up till now, we have used only very general features of scattering theory. 
In order to illustrate how microscopic details of the measurement device may result in domination of either of the transition amplitudes $b_\pm$, we construct a schematic model of a measurement device of stepwise increasing size. It is assumed that each step contributes 
a factor, close to $1$, depending on microscopic details $\alpha$ of the device, so that after $N$ steps we have a resulting product of $N$ independent factors. For simplicity, we assume that each factor is enhancing one channel and suppressing the other. Mathematically, this can be expressed as
\begin{align}
\label{b-amp}
& |b_+(\alpha)|^2 = \prod_{n=1}^N \big( 1 + \kappa_n(\alpha) \big) = 
e^{\Xi(Y(\alpha)-\tfrac{1}{2})} \;, \\
& |b_-(\alpha)|^2 = \prod_{n=1}^N \big( 1 - \kappa_n(\alpha) \big) = 
e^{\Xi(- Y(\alpha)-\tfrac{1}{2})} \;, \nonumber
\end{align}
where $\kappa_n(\alpha)^*=\kappa_n(\alpha)$, and where the small deviations from unity in the factors are characterised by $\langle\langle \kappa_n \rangle\rangle=0$, $\langle\langle \kappa_n\kappa_{n'} \rangle\rangle=\delta_{nn'}\kappa^2$, $0<\kappa<<1$, and $\Xi=N\kappa^2$. We have followed the convention to calculate to second order in $\kappa_n$ and then replace $\kappa_n \kappa_{n'}$ by its mean 
$\delta_{n n'}\kappa^2$. In (\ref{b-amp}), we have introduced the aggregate variable,
\begin{align}
\label{Y}
Y(\alpha)=\frac{1}{\Xi} \sum_{n=1}^N \kappa_n \; .
\end{align}
representing the overall degree of enhancement/suppression
(so that $Y>0$ for net enhancement of $+$ and $Y<0$ for net enhancement of $-$). The resulting squared amplitudes in (\ref{b-amp}) have the means unity, $\langle\langle |b_\pm|^2 \rangle\rangle = \langle\langle e^{\Xi(\pm Y-\tfrac{1}{2})} \rangle\rangle = 1$. An extra common factor for the amplitudes would not make any difference. For $Y$ in (\ref{Y}), the mean and variance are $\langle\langle Y \rangle\rangle=0$ and $\langle\langle Y^2 \rangle\rangle=\Xi^{-1}$.

For a large $\Xi$, i.e., for a sufficiently large number $N$, and a non-zero $Y$, the ratio of the squared moduli of the amplitudes $|b_+|^2/|b_-|^2=e^{2\Xi Y}$ becomes very large or very small. This demonstrates how the bifurcation can result from the unknown details of the initial state of the apparatus. Thus the condition for $\mu$ ending up in an eigenstate of $\sigma_3$ is fulfilled; our reasoning above showed that under the same condition, the eigenstates are reached with probabilities given by Born's rule.

In order to see all this more in detail for our model, we discuss the statistics of initial states and final states. Since all steps of extension in (\ref{b-amp}) are independent, the distribution   over the aggregate variable $Y$, defined by Equation (\ref{Y}), in the ensemble of initial states of $A$, is well described by the Gaussian distribution,
\begin{align}
\label{Gaussian}
q(Y)=\sqrt{\frac{\Xi}{2\pi}} \; e^{-\tfrac{1}{2}\Xi Y^2} \;.
\end{align}
The rate of the transition (\ref{transitionM}) with $|b_\pm|^2$ given by (\ref{b-amp}) is $(2 \pi)^{-1} w(Y)$ where 
\begin{equation}
\label{w(Y)}
w(Y)=|\psi_{+} |^2 e^{\Xi(Y-\tfrac{1}{2})} + | \psi_{-} |^2 e^{\Xi(- Y-\tfrac{1}{2})} \; ,
\end{equation}
with $\langle\langle w(Y) \rangle\rangle=1$. In (\ref{w(Y)}), the first term is the transition rate to the state $\ket{+}_\mu \otimes \ket{+,\beta_+(\alpha)}_A$ and the second term is the rate to the state $\ket{-}_\mu \otimes \ket{-,\beta_-(\alpha)}_A$.

\begin{figure}[htpb]
\begin{center}
\includegraphics[scale=.25]{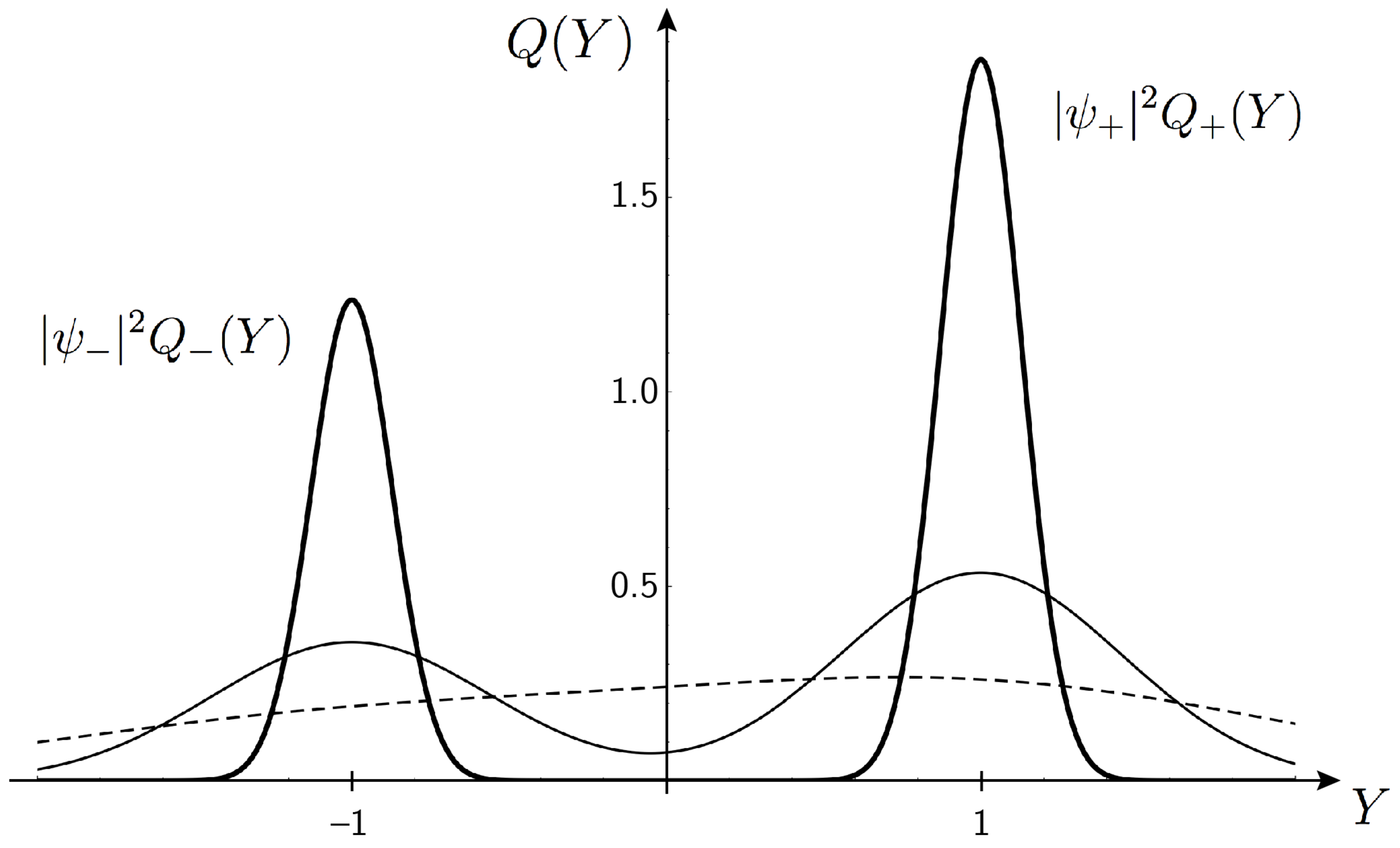}
\caption{The distribution $Q(Y)$ over $Y$ of transitions taking place in $\mu A$-interaction for increasing size of $A$ corresponding to $\Xi=1$ (broken line), $\Xi=5$ (thin line), and $\Xi=60$ (thick line). $Q(Y)$ is composed of 
two distributions $Q_+(Y)$ and $Q_-(Y)$
with weights $|\psi_{+} |^2$ and $|\psi_{-} |^2$, respectively. These distributions become separated as $\Xi$ increases. 
Each initial state of $A$, $\ket{0,\alpha}_A$, is represented by a certain $Y=Y(\alpha)$. As the size of the measurement device increases and $\Xi$ becomes larger, states that are efficient in leading to a transition are found around $Y=-1$ and $Y=+1$, respectively. These initial states then lead to $\mu$ ending up in either $\ket{-}_\mu$ or $\ket{+}_\mu$, respectively, with probabilities confirming the Born rule.}
\label{fig:Q-distribution}
\end{center}
\end{figure}

The total transition rate (\ref{w(Y)}) depends strongly on $Y$. We shall now go into the statistics of the final states which is strongly influenced by $w(Y)$. To get the distribution $Q(Y)$ over $Y$ for the {\it final states}, corresponding to $q(Y)$ for the initial states, we must multiply $q(Y)$ by the transition rate (\ref{w(Y)}) which is normalized in the sense that its mean value is $1$. This is the standard approach in scattering theory, see, e.g., Ref. \cite{Jauch1955}. Here, it can be interpreted as a selection process, as previously discussed, that favours initial states that are efficient in leading to a transition, with a selective fitness being proportional to the transition rate (\ref{w(Y)}). We then get (see Figure \ref{fig:Q-distribution})
\begin{align}
\label{Q(Y)}
&Q(Y)=q(Y) w(Y) = |\psi_{+} |^2 Q_+(Y) + | \psi_{-} |^2 Q_-(Y) \; , \nonumber \\
&Q_\pm(Y) = \sqrt{\frac{\Xi}{2\pi}} \; e^{-\tfrac{1}{2}\Xi(Y \mp 1)^2} \;.
\end{align}
The normalized partial distributions, $Q_+(Y)$ and $Q_-(Y)$, also with variance $\Xi^{-1}$, are centered around $Y=1$ and $Y=-1$ 
and correspond to $\mu$ ending up in the state $\ket{+}_\mu$ and $\ket{-}_\mu$, respectively. The coefficients of $Q_+(Y)$ and $Q_-(Y)$ in (Q(Y)), $|\psi_+|^2$ and $|\psi_-|^2$, express Born's rule explicitly.

It is instructive to follow the distribution $Q(Y)$ with growing $\Xi$. For small $\Xi$ ($=N\kappa^2$) it is broad and unimodal; it then turns broad and bimodal with narrowing peaks. For large $\Xi$, it is split into two well separated distributions with sharp peaks, 
weighted by the squared moduli of the state components of $\mu$,
$|\psi_{+} |^2 Q_+(Y)$ and $|\psi_{-} |^2 Q_-(Y)$, at $Y=1$ and $Y=-1$, respectively. They represent two different subensembles of final states (see Equation (\ref{final states})). Other values of $Y$ correspond to non-competitive processes. The aggregate variable $Y$ is "hidden" in the fine unknown details of $A$ that can influence the $\mu A$-interaction.

The initial state for $\mu$ in (\ref{initial state}) is a superposition, a '{\it both-and} state', and it ends up in (\ref{final states}) which is again a product state, with $\mu$ in {\it either} $\ket{+}_\mu$ {\it or} $\ket{-}_\mu$. The initial states of $A$ vary widely in their efficiency to lead to a final state. When one transition-rate term in (\ref{w(Y)}) is large, the other one is small. The selection of a large transition rate therefore also leads to a bifurcation with one of the terms in (\ref{transitionM}) totally dominating the final state.

{\it Generic model of $A$ and the $\mu A$-interaction.}---To give some hint of the physical meaning of our mathematical modelling, we sketch a generic model for $A$. We let $\mu$ be a fast incoming electron in a prepared spin state. To measure $\sigma_3$, we separate the $\ket{+}_\mu$ and the $\ket{-}_\mu$ components in an inhomogeneous magnetic field and send the $\ket{+}_\mu$ component into a detector where $\mu$ has a possibility to initiate a visible track by ionizing molecules along its path. The $\ket{-}_\mu$ component is led outside the detector and thus $b_-(\alpha)=1$. 
In this respect our example differs from the model of Equation (\ref{b-amp}).
A recorded detection means the result $+$; no detection means the result $-$.

We think of the component of $\mu$ passing through the detector as a small wavepacket; we let $A$ be a small cylinder of the detector material around the path of the wavepacket. We take successive small steps to build up $A$ by small increases of the cylinder diameter. Each unknown factor in the squared amplitude $|b_+|^2$, of Equation (\ref{b-amp}), coming from an extension of $A$, makes $|b_+|^2$ grow or shrink while $|b_-|^2=1$. Also in this case $\langle \langle |b_+|^2 \rangle \rangle=1$ expresses the metastability of $A$.

What starts as a state of $A$ with small components with ionized molecules along the path of $\mu$ can either develop into a state with sufficient seeds for condensation or boiling or relax back into a state of neutral molecules. An ambivalence in the selection of these alternatives is constitutional for the necessary metastability of the detector. The parameters $\kappa_n(\alpha)$ model this ambivalence in the transition amplitudes to a final state for $A$. These parameters depend on fine accidental details of its ingoing state $\ket{0,\alpha}_A$. 

{\it Discussion.}---In our description, we want the system $A$ to be big enough for a bifurcation to take place. We leave out the irreversible development beyond $A$. Our idea is to follow the qualitative recipe given by Bell who formulated a principle concerning the position of the Heisenberg cut \cite{Bell2004}, i.e., the boundary of the system $A$, interacting with $\mu$ according to quantum dynamics (Ref. \cite{Bell2004}, p.124):
\begin{quotation}
\noindent
{\it put sufficiently much into the quantum system that the inclusion of more would not significantly alter practical predictions}
\end{quotation}

In the model that we have described, the bifurcation of measurement takes place in the reversible stage of the interaction between $\mu$ and $A$ before irreversibility sets in and fixes the result. In this respect, our analysis is very different from 
decoherence analysis \cite{Zeh1970,Schlosshauer2005} and also from the approach in Zurek's work \cite{Zurek2003},
where the bifurcation is considered as a part of an irreversible process.

The system $A$ should not be so large that $\mu \cup A$ cannot be described by deterministic quantum dynamics. Still, it must be possible to have the entanglement process sufficiently extensive, i.e., to have $\Xi=N \kappa^2$ sufficiently large. Then we have followed Bell's principle quoted above concerning the position of the Heisenberg cut.

For future work, a more detailed description is needed of a typical $\mu A$-interaction, including the statistics of the initial states and the selection of one state $\ket{0,\alpha}_A$ with a large transition amplitude, leading to a final state (\ref{final states}) with $\mu$ in one eigenstate, $\ket{+}_\mu$ or $\ket{-}_\mu$. An important task is to construct a detailed physical model of a non-biased measurement apparatus like the one sketched above.

So far, the unitarity of the scattering matrix has not been explicitly visible. Reversibility that we have pointed out as crucial, 
is also not explicit. 
To remedy this we have made a slightly more elaborate description of the whole process where the observed system $\mu$ is produced in its initial state $\ket{\psi}_\mu$ by an external source before interaction with $A$ and absorbed by a sink in one of the possible final states after the interaction. Then both unitarity and reversibility 
are made explicit. The calculation has been done through evaluation of Feynman diagrams summed to all orders in perturbation theory. (This version is presented in the Supplemental Material \cite{supplementary}.)

We have also checked that it is straightforward to generalize our model to (i) any number of possible measurement results, and (ii) a system $A$ entangled with its environment and not possible to describe as being in a pure initial state.

If we had shown the development of $|b_\pm|^2$ for each step instead of going directly to the resulting product in Equation (\ref{b-amp}), in mathematical terms, we would have seen a quantum diffusion process close to the one described by Gisin and Percival \cite{Gisin1984,Gisin1992,Percival1994}. 

In earlier work, non-linearities have sometimes been brought in through generalization of quantum mechanics. Besides the quantum diffusion model, the Ghirardi-Rimini-Weber model is of this kind \cite{Ghirardi1986,Ghirardi1990}. In our model, we have seen how non-linearities can arise within quantum mechanics as higher-order terms in a perturbation expansion without any generalization.

In practical scientific research, there is a common working understanding of quantum mechanics. Physicists have a common reality concept for a quantum-mechanical system when it is not observed, a kind of pragmatic quantum ontology with the {\it quantum-mechanical state} of the studied system as the basic concept. Development of this state in time then constitutes the quantum dynamics. If quantum mechanics now can also be used to describe the measurement process, this pragmatic quantum ontology can have a wider validity than has been commonly expected.

Quantum mechanics deserves to be recognized as a realistic and deterministic theory and not only considered as a set of calculation rules with some spooky or weird features. A better understanding of quantum mechanics is essential at a time of fast progress both in experimental knowledge of quantum processes and in quantum technology.

\subsection*{Acknowledgements}
We thank Erik Sj\" oqvist and Martin Cederwall for fruitful collaboration in an earlier phase of this project \cite{Eriksson2017}. Financial support from The Royal Society of Arts and Sciences in Gothenburg was important for this collaboration. We are also grateful to Andrew Whitaker for several constructive discussions.

\normalem
\bibliographystyle{unsrt}
\bibliography{QM}

\begin{thebibliography}{10}
\bibitem{supplementary}
See {S}upplemental {M}aterial that includes a clarification of the role of the
  scattering and the transition matrices, a detailed description of the
  statistics of final states, a {F}eynman diagram representation of the model
  that makes inverse processes and unitarity explicit, and a number of relevant
  citations from the history of quantum measurement.
\end{thebibliography}

\end{document}


\section*{Supplemental material}
\renewcommand\theequation{{S.}\arabic{equation}}
\setcounter{equation}{0}
\renewcommand\thefigure{{S.}\arabic{figure}}
\setcounter{figure}{0}

\subsection*{S.1 Scattering matrix $S$ and transition matrix $M$}
\label{sec:scattering}

The unitary (i.e., probability preserving) scattering operator $S$, takes an initial state $\ket{i}$ into a final state $S\ket{i}$. If a certain final state $\ket{f}$ is of interest to us then we calculate the scattering-matrix element $\bra{f} S \ket{i}$. When dealing with particle scattering, it is convenient to do this in momentum space. Eigenstates of momentum are plane waves, i.e., states that occupy all space and cannot be normalized.

We shall be interested in final states $\ket{f}$ that are different from the initial state $\ket{i}$, so that $\ket{f}$ and $\ket{i}$ are orthogonal, i.e., $\langle f | i \rangle=0$, and we can replace $S$ by $S-1$.

We use here the Quantum Electrodynamics book by Jauch and Rohrlich as our reference \cite{Jauch1955}, to emphasize the development that had taken place between the physics of the 1930s and the quantum field theory of the 1950s.

To take into account energy and momentum conservation, it is usual to write (Ref. \cite{Jauch1955}, Eq. (8-29))
\begin{equation}
\label{S-1}
\bra{f} (S-1) \ket{i} = \delta (P_f - P_i) \bra{f} M \ket{i} \; ,
\end{equation}
where $\delta (P_f - P_i)$ is the 4-dimensional delta-function over energy-momentum and $M$ is the transition matrix.

Usually the probability for a transition into the final state $\ket{f}$, given the initial state $\ket{i}$, would be the squared modulus of (\ref{S-1}) but the square of a delta function does not make sense. Then one imposes a very large but finite length $L$ in space and requires normalization for the wave-functions in the volume $L^3$, and, similarly, one imposes a time $T$ for the whole process. Energy-momentum conservation is nearly exact for large $L$ and $T$. One delta function in the squared modified (\ref{S-1}) becomes replaced by $(2\pi)^{-4}L^3T$. When normalization conventions are taken into account, the result becomes independent of $L$ and proportional to $T$. After this we divide by $T$ to get the transition probability per unit time (see Ref. \cite{Jauch1955} (8-40)),
\begin{equation}
(2\pi)^{-1} \delta (P_f - P_i) | \bra{f} M \ket{i} |^2  \; .
\end{equation}
Then requesting the states $\ket{i}$ and $\ket{f}$ to have the same energy and momentum, we can interpret
\begin{equation}
(2\pi)^{-1} | \bra{f} M \ket{i} |^2 = (2\pi)^{-1} \text{Tr} [ \ket{f}\bra{f} M \rho^{(0)} M^\dagger ]  \; .
\end{equation}
as the transition probability per unit time, induced by $M$, from an initial state described by the density operator
\begin{equation}
\label{rho0}
\rho^{(0)} = \ket{i}\bra{i}
\end{equation}
to a final state described by the projection operator $ \ket{f}\bra{f}$.
We thus find that the transition probability-rate matrix obtained from the initial state (\ref{rho0}) is $(2\pi)^{-1}$ times
\begin{equation}
\label{R}
R= M \rho^{(0)} M^\dagger \; .
\end{equation}
Thus $(2\pi)^{-1} R$ is the total transition rate times the density operator for the final state. Since the trace of a density operator is unity,
\begin{equation}
(2\pi)^{-1}  w = (2\pi)^{-1}\, \text{Tr} \, R 
\end{equation}
is the total transition rate. The normalized final-state density matrix is then
\begin{equation}
\rho^{(f)} = \frac{1}{w} R = \frac{M \rho^{(0)}M^\dagger}{\text{Tr}[M \rho^{(0)}M^\dagger]} \; .
\end{equation}
Let us consider the systems $\mu$ and $A$ of the article. Let $M$ make $A$ entangled with $\mu$ without changing the state of $\mu$. Still the transition amplitudes can differ between $+$ and $-$. This can distort the entanglement and induce changes in the relative proportions of $+$ and $-$ in the final state (Equation (4) in the article). Thus the proportions are no longer fixed by the von Neumann dilemma; the dilemma does not arise in the scattering theory that we are considering.

\subsection*{S.2 Statistics of transitions to a final state}
We consider initial states of the system $A$ with a density matrix of the form
\begin{equation}
\label{init-dens-mtrx}
\ket{0,\alpha}{}_A \;{}_A\bra{0,\alpha} \; .
\end{equation}
The density matrix for the initial state of the combined system $\mu \cup A$ is then
\begin{align}
\label{comb-init-dens-mtrx}
\rho^{(0)}(\alpha)=\ket{\psi}{}_\mu \; {}_\mu \bra{\psi} \otimes \ket{0,\alpha}{}_A \;{}_A\bra{0,\alpha} \; .
\end{align}
It corresponds to the (non-normalized) final state density matrix (\ref{R}),
\begin{align}
\label{final-dens-mtrx}
&R(\alpha) = M \rho^{(0)}(\alpha) M^\dagger = 
M \Big( \ket{\psi}{}_\mu \; {}_\mu \bra{\psi} \otimes \ket{0,\alpha}{}_A \;{}_A\bra{0,\alpha} \Big)M^\dagger = \nonumber \\
&\;\; \;\; = \sum_{j,k=\pm} R_{jk}(\alpha) \ket{j}{}_\mu \; {}_\mu \bra{k} \otimes \ket{j, \beta_j(\alpha)}{}_A \; {}_A\bra{k, \beta_k(\alpha)} \;; \\
&R_{jk}(\alpha) = b_j(\alpha)b_k(\alpha)\psi_j \psi_k^*   \; .\nonumber
\end{align}
Here $b_\pm(\alpha)$ are the scattering amplitudes. In our model they are given by Equation (4) of the article,
\begin{equation}
\label{b-eq}
b_\pm(\alpha)=e^{\tfrac{1}{2}\Xi\big(\pm Y(\alpha)-\tfrac{1}{2}\big)} \; ,
\end{equation}
where $Y(\alpha)$ is an aggregate variable for the unknown enhancement/suppression factor of $A$, and $\Xi = N \kappa^2$. The product of the amplitudes (\ref{b-eq}) is 
\begin{equation}
\label{b+-}
b_+(\alpha) b_-(\alpha) = e^{-\tfrac{1}{2}\Xi} \;.
\end{equation}
The means of the squared amplitudes are
\begin{align}
\label{b-mean}
\big\langle\big\langle |b_\pm(\alpha)|^2 \big\rangle\big\rangle = 
\Big\langle\Big\langle e^{\Xi(\pm Y -\tfrac{1}{2})} \Big\rangle\Big\rangle = 1 \;.
\end{align}
The trace of $R(\alpha)$ is the total transition rate (apart from a factor $(2\pi)^{-1}$),
\begin{align}
\label{tot-rate}
&\text{Tr} R(\alpha) = |b_+(\alpha)|^2 |\psi_+|^2+|b_-(\alpha)|^2 |\psi_-|^2 = w(Y(\alpha)) \;,\nonumber \\
&w(Y) = e^{\Xi(Y-\tfrac{1}{2})}|\psi_+|^2 + e^{\Xi(-Y-\tfrac{1}{2})}|\psi_-|^2    \; ;  \\[4pt]
&\langle\langle w(Y) \rangle\rangle = 1   \;.\nonumber 
\end{align}
The matrix $R(\alpha)$ in (\ref{final-dens-mtrx}) still describes a pure state,
\begin{align}
\sum_{k=\pm} R_{jk}(\alpha) R_{kl}(\alpha) = w(\alpha) R_{jl}(\alpha) \;.
\end{align}
Equation (\ref{b+-}) tells us that the non-diagonal elements of $R(\alpha)$ become very small for large $\Xi$. 
The amplitudes (\ref{b-eq}) are functions of $\alpha$ only through the aggregate variable $Y(\alpha)$. 

Let the ensemble of available ingoing states for $A$ be described by the density operator
\begin{equation}
\sum_\alpha \nu_\alpha \ket{0,\alpha}{}_A\;{}_A\bra{0,\alpha} \; , \;\; \sum_\alpha \nu_\alpha = 1 \;.
\end{equation}
The measure of the region in $\alpha$-space having $Y(\alpha)=Y$, is (per unit length of $Y$)
\begin{align}
q(Y) = \sum_\alpha \nu_\alpha \, \delta (Y(\alpha)-Y) \;; \;\;\;   \int_{-\infty}^\infty d Y q(Y) = 1 \;.
\end{align}
We have for $Y$ the mean and variance
\begin{align}
&\int_{-\infty}^\infty d Y q(Y)Y = \sum_\alpha \nu_\alpha Y(\alpha) =  \langle\langle Y \rangle\rangle = 0 \;; \nonumber \\
&\int_{-\infty}^\infty d Y q(Y)Y^2 = \sum_\alpha \nu_\alpha Y(\alpha)^2 = \langle\langle Y^2 \rangle\rangle = \frac{1}{\Xi} \;. 
\end{align}
Since $Y$ is composed of many small contributions, it is reasonable to equate the distribution $q(Y)$ with a Gaussian (Equation (6) in the article),
\begin{align}
\label{Gaussian2}
q(Y)=\sqrt{\frac{\Xi}{2\pi}} \; e^{-\tfrac{1}{2}\Xi Y^2} \;.
\end{align}
The squared amplitude $|b_j(\alpha)|^2 \; (j=+,-)$ satisfying (\ref{b-mean}) is proportional to the probability density for a final state with $\mu$ in the state $\ket{j}{}_\mu \;{}_\mu\bra{j}$ to have come from an initial state $\ket{j}{}_\mu \;{}_\mu\bra{j} \otimes \ket{0,\alpha}{}_A \;{}_A\bra{0,\alpha}$. The probability density for a final state with $\mu$ in the state $\ket{j}{}_\mu \;{}_\mu\bra{j}$ to have come from an initial state $\ket{j}{}_\mu \;{}_\mu\bra{j} \otimes \ket{0,\alpha}{}_A \;{}_A\bra{0,\alpha}$ for which $Y(\alpha)=Y$, is then
\begin{align}
&Q_\pm(Y) = \sum_\alpha \nu_\alpha \delta(Y(\alpha)-Y) |b_\pm(\alpha)|^2 = q(Y) e^{\Xi(\pm Y-\tfrac{1}{2})} 
= \sqrt{\frac{\Xi}{2\pi}} e^{-\tfrac{1}{2}\Xi(Y\mp 1)^2} \;;  \\
&\int_{-\infty}^\infty dY \, Q_\pm(Y) = 1 \;,\nonumber
\end{align}
The total distribution of final states over $Y$, starting from $\rho^{(0)}(\alpha)$ in (\ref{comb-init-dens-mtrx}), is
\begin{align}
\label{S.26}
&Q(Y) = q(Y) \Big( |\psi_+|^2 e^{\Xi(Y-\tfrac{1}{2})} +   |\psi_-|^2 e^{\Xi(-Y-\tfrac{1}{2})} \Big) =
|\psi_+|^2 Q_+(Y) +   |\psi_-|^2 Q_-(Y) \;.
\end{align}
This gives some background for the distributions in Equation (8) in the article.

\subsection*{S.3 Calculation of Feynman diagrams with explicit unitarity and reversibility}
The unitarity of the scattering matrix has not been explicitly visible in the main text. Reversibility that we have pointed out as crucial, is also not explicit. To remedy this we shall present a slightly more elaborate description of the whole process where the observed system $\mu$ is produced in its initial state $\ket{\psi}_\mu$ by an external source $B$ before interaction with $A$ and absorbed by a sink $D_+$ or $D_-$ in one of the possible final states after the interaction. In this version both unitarity and reversibility will be made explicit.

In this picture, the transition rate will instead be hidden and hence also the race to the final state. We therefore use the results that we have already obtained in the article, the transition rate (7) and the distribution (8) of the final states over the aggregated variable $Y$. The Born rule is also contained in (8).

As in the previous description, $A$ starts in the initial state $\ket{0,\alpha}_A$ but $\mu$ is produced by $B$ at an early time $-T$ in the state $\ket{\psi}_\mu$. After $\mu A$-interaction around the time zero, $\mu$ is absorbed in an eigenstate $\ket{+}_\mu$or $\ket{-}_\mu$ at the time $+T$ by $D_+$ or $D_-$, leaving $A$ in the state $\ket{+,\beta_+(\alpha)}_A$ or $\ket{-,\beta_-(\alpha)}_A$, respectively. We thus have one initial state $\ket{0,\alpha}_A$, a member of the ensemble of initial states, and three available final states, $\ket{0,\alpha}_A$ (no change), $\ket{+,\beta_+(\alpha)}_A$, and $\ket{-,\beta_-(\alpha)}_A$; The system $\mu$ takes part only in intermediate states.

Feynman-diagram elements for the action of the source $B$, the transition matrix $M$ in Equation (3) of the article and the sinks $D_+$ and $D_-$ are shown in Figure S.1, and the factors corresponding to them, $J^*$, $b_\pm \psi_\pm$ and $F_\pm$. We represent $\mu$ by a thin line and $A$ by a thick line. As in Figure 1 of the article, the interaction between $\mu$ and $A$ described by the transition matrix $M$, is represented by a shaded circle. Reversibility is included through the actions of the hermitean or complex conjugates, $J$, $M^\dagger$, and $F_j^*$.

\begin{figure}[htpb]
\begin{center}
\includegraphics[scale=1.]{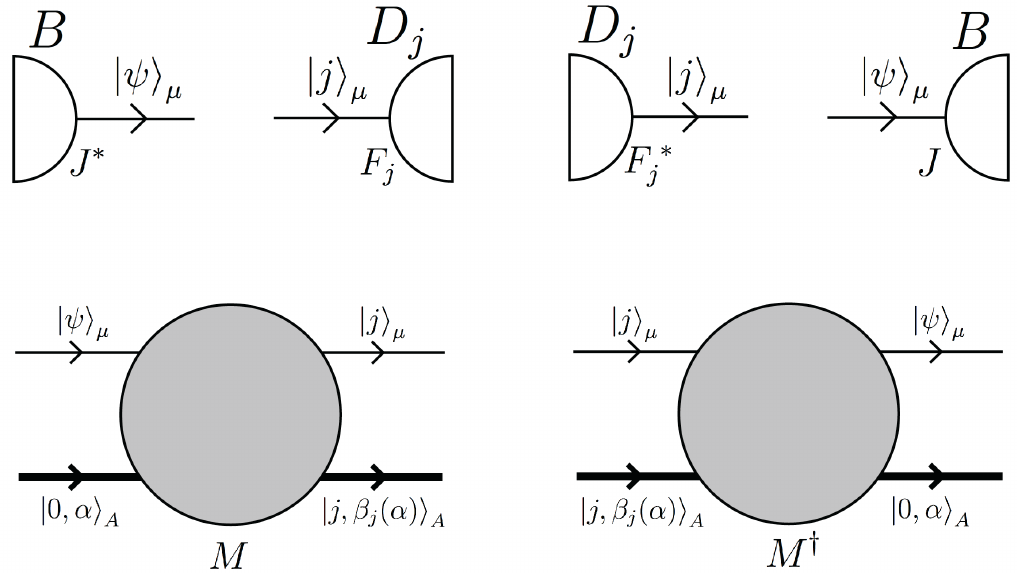}
\caption{Feynman-diagram elements for the action of the source $B$, the transition matrix $M$ and the sinks $D_j \;(j=\pm)$ and their conjugates.}
\label{fig:S.1}
\end{center}
\end{figure}

\begin{figure}[htpb]
\begin{center}
\includegraphics[scale=0.9]{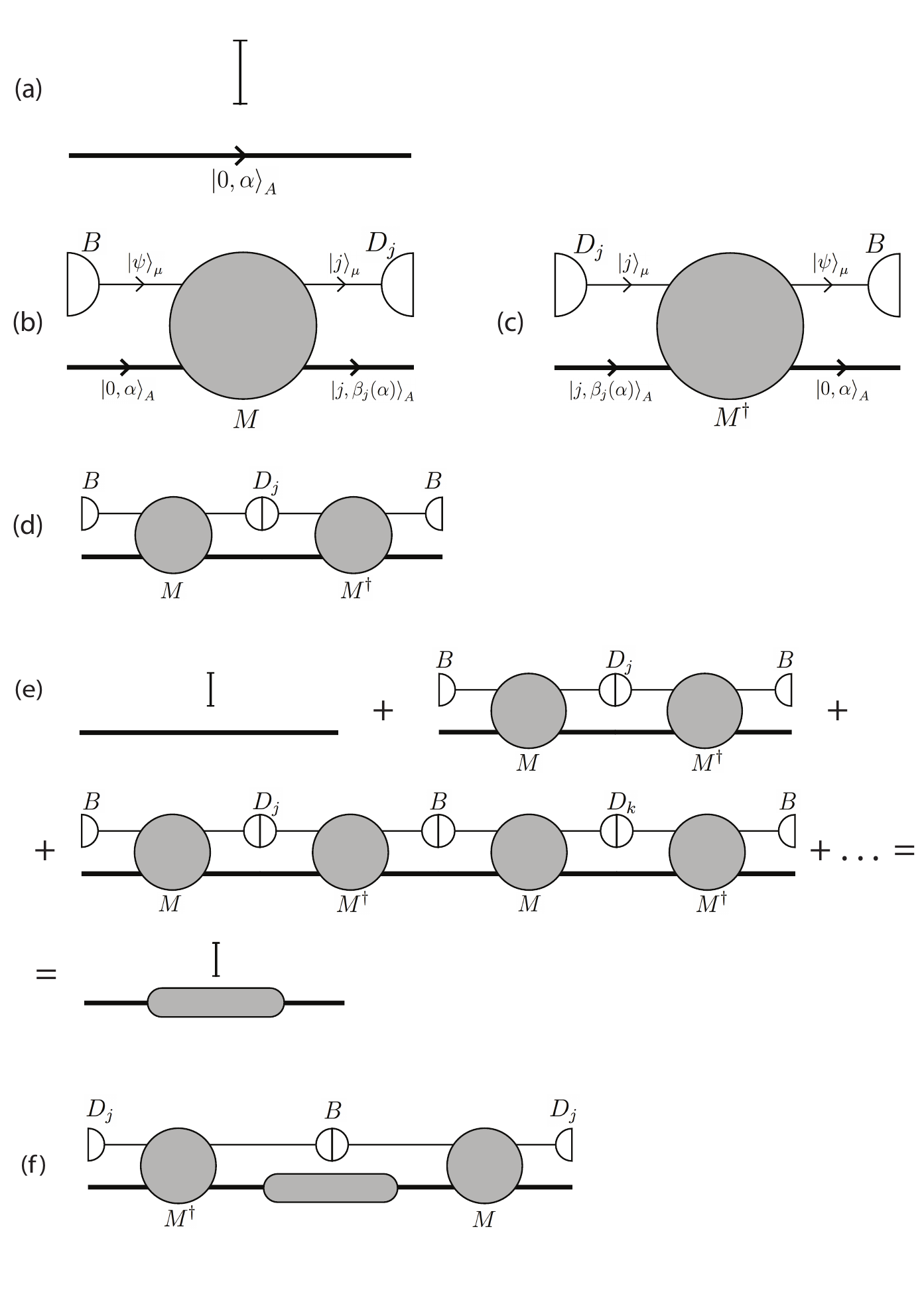}
\caption{Feynman diagrams: (a) zero order diagram for no change (the I-like sign above the $A$-line symbolizes "no $\mu$-system"); (b) lowest order diagram for transition to a state with $A$ marked by $\mu$ in the $j$ state (see Fig. 1 in the article); (c) inverse diagram of b; (d) diagrams b and c combined to a no-change correction; (e) summation over d repeated any number of times, i.e., summation of no-change diagrams to all orders; (f) the full perturbation expansion of the diagonal elements of the final-state density matrix with $A$ marked by $\mu$ in the state $\ket{j}_\mu$.}
\label{fig:S.2}
\end{center}
\end{figure}

We use perturbation theory to compute the final-state density matrix,
\begin{align}
S \ket{0,\alpha}{}_A \;{}_A\bra{0,\alpha} S^\dagger \; .
\end{align}
We use the method of Nakanishi \cite{Nakanishi1958} to calculate this bilinear quantity directly rather than the state vector $S \ket{0,\alpha}_A$, simply because it makes normalization easy.

The diagrams of perturbation theory are shown in Figure S.2. The zero-order no-change term is only an $A$-line corresponding to a contribution equal to 1 (Figure S.2a). Figure S.2b shows the diagram corresponding to that of Figure 1 of the article with the source $B$ and one sink $D_j  \;\; (j=+,-)$. The inverse of this diagram is that of Figure S.2c. The two taken together into one diagram represents a reduction of the no-change component due to transitions to the other states (Figure S.2d). This can be repeated any number of times. All these diagrams leading back to the initial state (Figure S.2e) contribute a geometrical series, representing the total no-change component of the final state,
\begin{align}
\label{no-change}
&1 - \sum_{j=+,-}  J\, \psi_j^* b_j F_j^* F_j b_j \psi_j J^*  +  
     \Big( \sum_{j=+,-} J\, \psi_j^* b_j F_j^* F_j b_j \psi_j J^* \Big)^2  \pm ... =  \\
&\frac{1}{1 + \Big( {|F_+|}^2 {|\psi_+|}^2|b_+|^2 + {|F_-|}^2 {|\psi_-|}^2|b_-|^2 \Big) {|J|}^2} =
\frac{1}{1+{|J|}^2 {|F|}^2 \Big(  {|\psi_+|}^2 e^{\Xi(Y-\tfrac{1}{2})}  +  {|\psi_-|}^2 e^{\Xi(-Y-\tfrac{1}{2})}  \Big)} \nonumber \;.
\end{align}

Here we have used the expressions for the amplitudes in Equation (4) of the article and given equal strength $F$ to the two sinks $D_+$ and $D_-$. The total scattering probability, i.e., the probability of $A$ being marked by $\mu$ is
\begin{align}
\label{scattering-prob}
&1 - \frac{1}{1+{|J|}^2 {|F|}^2 \Big(  {|\psi_+|}^2 e^{\Xi(Y-\tfrac{1}{2})}  +  
{|\psi_-|}^2 e^{\Xi(-Y-\tfrac{1}{2})}  \Big)} \nonumber = \nonumber \\
& J {\psi_+}^* e^{\tfrac{1}{2}\Xi(Y-\tfrac{1}{2})} F^* \frac{1}{1+{|J|}^2 {|F|}^2 
\Big(  {|\psi_+|}^2 e^{\Xi(Y-\tfrac{1}{2})} + {|\psi_-|}^2 e^{\Xi(-Y-\tfrac{1}{2})}  \Big)}
F e^{\tfrac{1}{2}\Xi(Y-\tfrac{1}{2})} \psi_+ J^*  \;+ \\
& J {\psi_-}^* e^{\tfrac{1}{2}\Xi(-Y-\tfrac{1}{2})} F^* \frac{1}{1+{|J|}^2 {|F|}^2 
\Big(  {|\psi_+|}^2 e^{\Xi(Y-\tfrac{1}{2})} + {|\psi_-|}^2 e^{\Xi(-Y-\tfrac{1}{2})}  \Big)}
F e^{\tfrac{1}{2}\Xi(-Y-\tfrac{1}{2})} \psi_- J^*  \; . \nonumber
\end{align}
The two terms on the right side of (\ref{scattering-prob}) are the probabilities for the final states $\ket{+,\beta_+(\alpha)}_A$ and $\ket{-,\beta_-(\alpha)}_A$,
corresponding to the diagrams of Figure S.2f for the remaining diagonal elements of the density matrix. For large $\Xi$, the no-change contribution (\ref{no-change}) becomes negligible. The same is true for the non-diagonal elements of the density matrix. The
diagonal terms for $+$ and $-$ in (\ref{scattering-prob}) become
\begin{align}
p_\pm = \frac{|\psi_\pm|{}^2 e^{\pm \Xi Y}}{|\psi_+|{}^2 e^{\Xi Y} + |\psi_-|{}^2 e^{-\Xi Y}} \;.
\end{align}
For $Y=+1$, $p_+=1$ and the $+$ channel takes everything and for $Y=-1$, $p_-=1$ and the $-$ channel takes everything. The norm is preserved, i.e., $S$ is unitary. Reversibility is also clearly visible: $J^*$, $M$ and $F_\pm$ are active together with their conjugates that represent inverse processes.

\subsection*{S.4 Quotations on the measurement problem}
Much has been written on the measurement problem during its long history. We give here a set of quotations that have been of special importance to us in different ways.

Richard Feynman was dissatisfied with the lack of understanding of measurement as a physical process. In The Feynman Lectures \cite{Feynman1963} he expressed this clearly:

\begin{quote}
[P]hysics has given up on the problem of trying to predict exactly what will happen in a definite circumstance. Yes! physics has given up. {\it We do not know how to predict what would happen in a given circumstance}, and we believe now that it is impossible, that the only thing that can be predicted is the probability of different events. It must be recognized that this is a retrenchment in our earlier ideal of understanding nature. It may be a backward step, but no one has seen a way to avoid it.
\end{quote}

\noindent
The uncertainty about what happens in a measurement was a feature of quantum mechanics from the very beginning. Niels Bohr attributed the situation to Nature itself \cite{Bohr1957}:

\begin{quote}
Step by step, we have been increasingly forced to refrain from describing the situation of single atoms in time and space with reference to the causal law and instead accept that nature has a free choice between different possibilities. The outcome of the choice, we can only predict probabilistically.
\end{quote}

\noindent
Eugene Wigner tried to base the reality of the external world on human consciousness \cite{Wigner1961}:

\begin{quote}
It may be premature to believe that the present philosophy of quantum mechanics will remain a permanent feature of future physical theories, [but] it will remain remarkable, in whatever way our future concepts may develop, that the very study of the external world led to the conclusion that the content of consciousness is an ultimate reality.
\end{quote}

\noindent
Erwin Schr\" odinger's famous Gedankenexperiment described a cat that was in a superposition of being alive in one component of its state and dead in another. Schr\" odinger's ambition was to show the absurdity of quantum mechanics but instead he opened a door for a new kind of speculative ideas concerning superpositions of macroscopically different states. Instead of trying to explain the 'reduction of the wave-function', one can stay with the von Neumann dilemma and deny that a reduction ever takes place. One then considers the quantum-mechanical time development to describe a wider reality.

Steven Weinberg described the relative-state interpretation  and its continuation in the many-worlds interpretation in this way as a consequence of the von Neumann dilemma \cite{Weinberg2017}:

\begin{quote}
[...] in consequence of their interaction during measurement, the wave function becomes a superposition of two terms, in one of which the electron spin is positive and everyone in the world who looks into it thinks it is positive, and in the other the spin is negative and everyone thinks it is negative. Since in each term of the wave function everyone shares the belief that the spin has one definite sign, the existence of the superposition is undetectible. In effect the history of the world has split into two streams, uncorrelated with each other.

\noindent
This is strange enough, but the fission of history would not only occur when someone measures a spin. In the realist approach the history of the world is endlessly splitting; it does so every time a macroscopic body becomes tied with a choice of quantum states. This inconceivably huge variety of histories has provided material for science fiction, and it offers a rationale for a multiverse[.]
\end{quote}

\noindent
Bryce DeWitt \cite{DeWitt1970} introduced the many-worlds interpretation in Physics Today as follows:

\begin{quote}
every quantum transition taking place on every star, in every galaxy, in every remote corner of the universe is splitting our local world on earth into myriads of copies of itself.
\end{quote}

\noindent
But he immediately hesitated:

\begin{quote}
I still recall vividly the shock I experienced on first encountering this multiworld concept. The idea of $10^{100+}$ slightly imperfect copies of oneself all constantly splitting into further copies, which ultimately become unrecognizable, is not easy to reconcile with common sense.
\end{quote}

\noindent
Like Feynman, Weinberg is not satisfied with the situation; he would prefer a one-world theory; in the quoted article \cite{Weinberg2017} he writes:

\begin{quote}
But the vista of all these parallel histories is deeply unsettling, and like many other physicists I would prefer a single history.
\end{quote}

\noindent
The attempt to understand measurement within quantum mechanics can be viewed as a consistency check. John Bell wrote the following about requesting a better understanding (Ref. \cite{Bell2004}, p. 125):

\begin{quote}
... the notion of the 'real' truth as distinct from the truth that is presently good enough for us, has also played a positive role in the history of science. Thus Copernicus found a more intelligible pattern by placing the sun rather than the earth at the centre of the solar system. I can well imagine a future phase in which it happens again, in which the world becomes more intelligible to human beings, including theoretical physicists, when they do not imagine themselves to be at the centre of it.
\end{quote}

\noindent
The reasons to search for a better understanding were very well expressed by Brian Greene \cite{Greene2005}:

\begin{quote}
[...] even though decoherence suppresses quantum interference and thereby coaxes weird quantum probabilities to be like familiar classical counterparts, {\it each of the potential outcomes embodied in the wavefunction still vies for realization}. And so we are still wondering how one outcome  "wins" and where the many other possibilities "go" when that actually happens. When a coin is tossed, classical physics gives an answer to the analogous question. It says that if you examine the way the coin is set spinning with adequate precision, you can, in principle, {\it predict} whether it will land heads or tails. On closer inspection, then, precisely one outcome is determined by the details you initially overlooked. The same cannot be said in quantum physics. Decoherence allows quantum probabilities to be interpreted much like the classical ones, but does not provide any finer details that select one of the many possible outcomes to actually happen.

Much in the spirit of Bohr, some physicists believe that searching for such an explanation of how a single, definite outcome arises is misguided. These physicists argue that quantum mechanics, with its updating to include decoherence, is a sharply formulated theory whose predictions account for the behavior of laboratory measuring devices. And according to this view, that is the goal of science. To seek an explanation of {\it what's really going on}, to strive for an understanding of {\it how a particular outcome came to be}, to hunt for a level of {\it reality beyond detector readings and computer printouts} betrays an unreasonable intellectual greediness.

Many others, including me, have a different perspective. Explaining data is what science is about. But many physicists believe that science is also about embracing the theories data confirms and going further by using them to get maximal insight into the nature of reality. I strongly suspect that there is much insight to be gained by pushing onward toward a complete solution of the measurement problem.
\end{quote}

\normalem
\bibliographystyle{unsrt}
\bibliography{QM}